# Open Source in Lab Management

Julien Cohen-Adad[1,2,3,4]

[1] *NeuroPoly Lab, Institute of Biomedical Engineering, Polytechnique Montreal, Montreal, QC, Canada*
[2] *Functional Neuroimaging Unit, CRIUGM, University of Montreal, Montreal, QC, Canada*
[3] *Mila - Quebec AI Institute, Montreal, QC, Canada*
[4] *Centre de recherche du CHU Sainte-Justine, Université de Montréal, Montreal, QC, Canada*

This document describes how managing a scientific lab can benefit from open source software and practices. It is not exhaustive, and very much biased towards particular software, some which have been chosen for historical reasons. One motivation for open source practices in science is to promote reproducibility and avoid situations like the one illustrated on **Figure 1**, that ultimately slows down scientific progress.

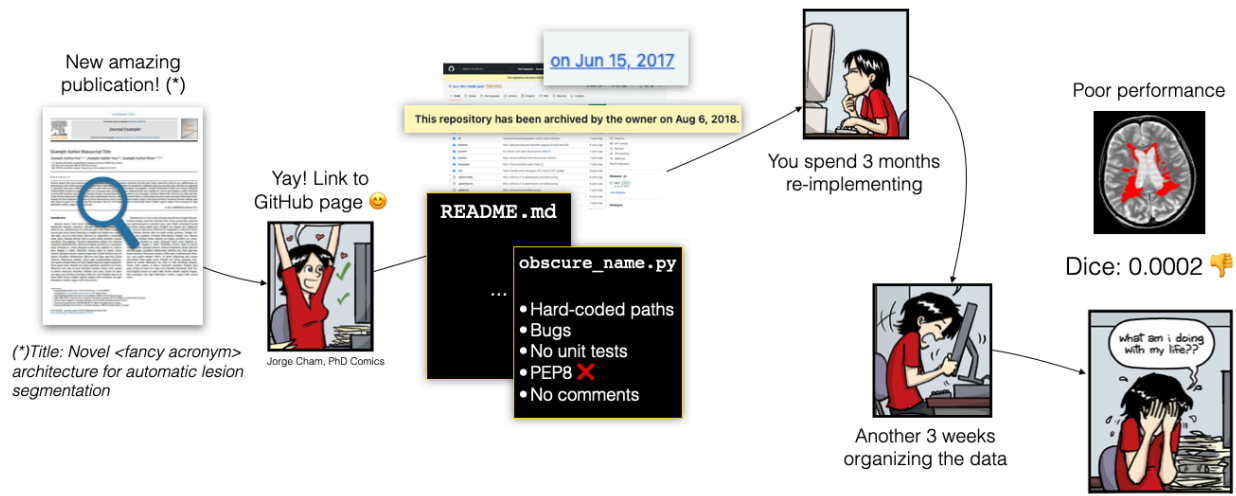

**Figure 1**. **The cycle of un-reproducible science**. You are a student working on the development of medical image analysis methods. You find a new paper with a promising algorithm. You notice the code is available on GitHub: Hurray! But wait… after looking into it, you realize the code is poorly documented, has bugs, hard-coded paths, etc. and it has been archived and is no longer maintained. The author of the repository has since left academia to get a job in industry and they have no interest in answering your question. That's fine. You know how to code. You spend time re-implementing the method, then you spend additional time organizing the messy data your collaborators sent you. Finally! You can test the method on your data. You prepare a bottle of champagne in the fridge. You run your script, come back 48h later to see the results: that method did not work. You can spend additional weeks/months tweaking the parameters, fixing the bugs, or you can also test one of the 1,000+ methods that are being published every year. Good luck. Some of the drawings are borrowed from PhD Comics (https://phdcomics.com/).

**Table of content**:

# Website

Every lab should have a website that describes the research performed in the lab, the team, publications, available positions, etc. Also useful is an intranet, which contains information useful for lab members, such as: onboarding procedures, relevant university courses, poster templates, procedure for conference reimbursement, who to acknowledge on papers/presentations, etc. At NeuroPoly we have a lab website: www.neuro.polymtl.ca and an intranet https://intranet.neuro.polymtl.ca/. They are both hosted on GitHub pages[1]. The source of the websites is on public GitHub repositories, is written in standard and easy to learn markdown format, and is readable directly from the GitHub web interface. Compilation of the website is done every time there is a change on the source (ie: someone 'git commit/push' a change) to generate the prettier, more public-facing version, which includes some extra document features like tabbed subpanes and sortable tables. The website can also be built locally to check if changes are OK before pushing. For important changes, we use pull requests. Everyone with a GitHub account can update the website via a pencil icon on every public-facing page (for example, to fix stale information, fix wrong links, add their information on the 'team' page). In addition to being open and flexible to manage, this approach is also an opportunity for new students in the lab to get familiar with git/github.

**Managing publications**: An important aspect of a lab's website is the management of publications. Manually maintaining the list of publications is prone to error and requires time. With the website hosted on GitHub and automatically compiled, an attractive solution is to leverage code that fetches the list of publications from a centralized source organized by the PI (eg: could be a google sheet), formats the publication list as a markdown page, and update that page on the lab's website source every time someone commits something on the website via a

---

[1] https://pages.github.com/

Github's Action workflow. As an example, the NeuroPoly publication list[2] is automatically updated via this github action[3] and the bibeasy software developed by NeuroPoly [1]. See an illustration of the workflow in **Figure 2**. What's interesting is the possibility to tweak the page (with javascript code) by adding e.g. filters to the publication based on keywords.

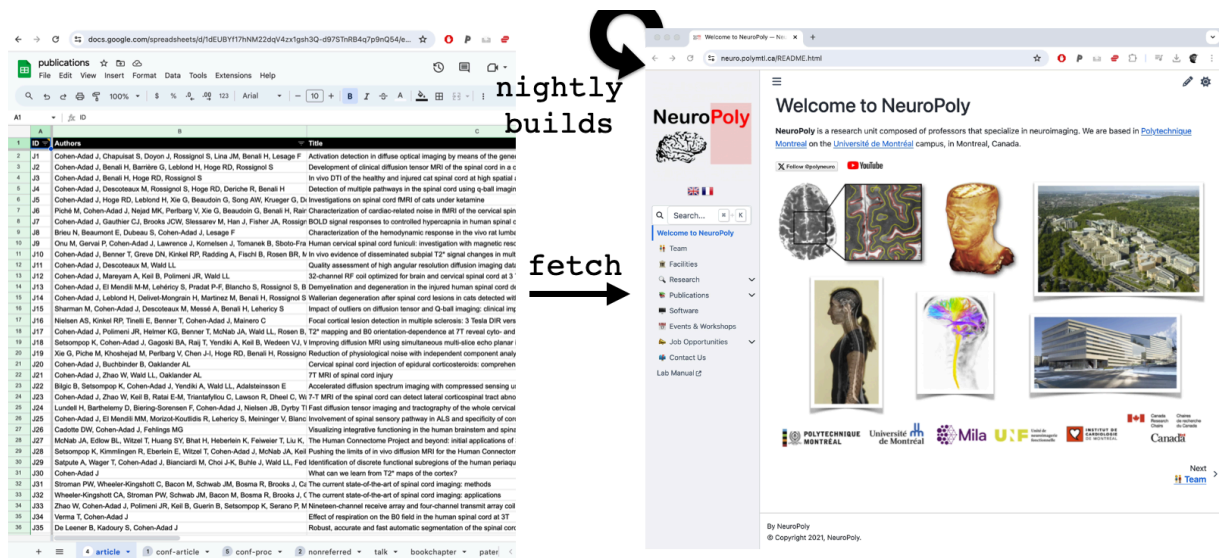

**Figure 2**. Automation of publication list generation for a lab's website using GitHub Actions.

## Organizing Datasets

**Standardizing file structure with BIDS:** We organize our public and private datasets according to the BIDS standard [2]. Procedure to BIDSify our datasets are on our intranet[4]. We created this custom doc instead of redirecting to the BIDS specs, because the latter is not 'easy' to understand for new students. So it is a more practical manual, with templates for creating new BIDS dataset and useful links. Esp: specific info about derivatives folder (eg: segmentation labels). BIDS also allows tracking data and label provenance using JSON sidecars containing information on which tool (eg: deep learning model) was used to generate the label and whether the label was manually corrected or not.

**Continuous testing of data compliance**: How can you ensure your data are and *stay* BIDS compliant? Use PyBIDS [3] combined with cron jobs such as Github Action (for cloud hosting), that runs every time new data is pushed to the repository. Can tests for missing file, acquisition parameters, etc. See this footnote[5] for an example of github action for continuous BIDS compliance testing.

**Hosting data**: Physical data are on OpenNeuro or on Amazon S3 servers for public data, and university servers (with backup routine) for private ones. Data are versioned with git-annex, which is built around git. The concept of git-annex is that instead of versioning large binary files (repos would grow very fast, not scalable), it versions the SHA256 of the pointer to the data. Some might be familiar with Datalad [4]: it is built on top of git-annex and adds functions for 'easier' UX. We decided to go with git-annex directly, to remove a software layer to have a better understanding of what each command does. Procedure to use our dataset with git-annex is in our intranet[6].

---

[2] https://neuro.polymtl.ca/publications/journal_articles.html
[3] https://github.com/neuropoly/neuro.polymtl.ca/blob/master/.github/workflows/publish.yml
[4] https://intranet.neuro.polymtl.ca/data/dataset-curation.html
[5] https://github.com/spine-generic/data-multi-subject/actions/runs/6203717476/workflow

The great thing with managing lab dataset with git/git-annex is that:

- When using the dataset, students can refer to the exact version that was used;
- If multiple students work on the same dataset, it ensures they are always in sync (eg: scenario where one student manually corrects a ground truth label to train a segmentation model, and the other student still uses the previous version of the label);
- It allows to conveniently add new data (e.g., from a new time point) while tracking the data provenance (who added the data and when);
- It ensures reproducible analyses; whereas data are private or public, the code to do the analysis (which itself if also versioned) will download the specific version of the data (the version is in the analysis script), and run the analysis. Code could be a SHELL, python script, or Jupyter Notebook, or other. Example in this footnote[7];

With git-annex/Datalad becoming the new trend in managing open access neuroimaging data, for transparent and reproducible analysis (e.g. YODA framework[8]), there are benefits in having students being exposed to these complex technologies, so they understand them better and are more prone to using them.

## IT management

**Configuration management, application deployment, orchestration**: Management of lab computers is a necessary pain. A very 'hands off' approach is to give students a laptop and have them install whatever they need. This approach works relatively well, until the laptop is returned and reused and needs to be reconfigured for the next student, or until there is a need to deploy shared resources (eg: CPU/GPU clusters, data server, shared printer, etc.), or proprietary software with network licenses (eg: Matlab, CST, Ansys HFSS). Also, not every PI has the budget to buy every student a laptop, and instead relies on desktops that stay in the lab, and that require centralized maintenance. An efficient way to manage servers and desktops is to use configuration management software such as Ansible[9], with which one can deploy a collection of software and configuration settings to a specific set of servers inside the lab's network. That way, when a new server is purchased, it is 'simply' added to the configuration script of Ansible, and all the necessary OS and software are installed automatically during the next deployment cycle, which can be scheduled automatically or run manually. This helps ensure reproducibility of scientific results, making sure that, in addition to the code and the data, a 'clean' OS and dependent software can be re-installed on demand. To improve reproducibility even further, systems should be wiped and reinstalled, and software upgraded on a predictable schedule. Automatic updates should be enabled on all systems that have them, and major OS version upgrades, which usually require manual intervention, should be planned and done in tight time windows, all to keep everyone's software in sync as much as possible.

---

[6] https://intranet.neuro.polymtl.ca/data/git-datasets.html
[7] https://github.com/shimming-toolbox/rf-shimming-7t
[8] https://handbook.datalad.org/en/latest/basics/101-127-yoda.html
[9] https://www.ansible.com/

**Monitoring stations**: Netdata[10] is highly effective for real-time system monitoring, offering detailed insights into various performance metrics like CPU usage and network traffic. It supports extensive integrations, customizable alerts, and both real-time and historical data analysis. **Figure 3** shows a screenshot of Netdata monitoring for our lab. Deploying Netdata using Ansible automates and standardizes monitoring across multiple nodes, streamlines updates, and integrates security practices, enhancing both efficiency and consistency in infrastructure management.

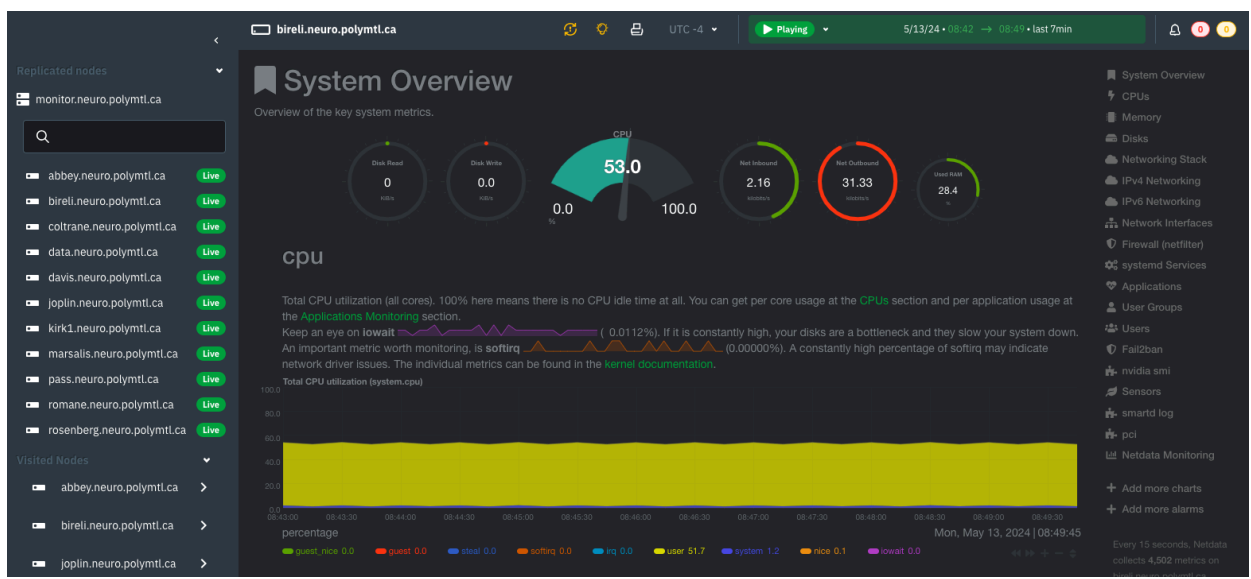

**Figure 3**. Monitoring of lab's computers hardware and network traffic using Netdata.

**Managing passwords:** Pass[11], the standard Unix password manager, offers a secure method for managing passwords within a lab setting. It utilizes GPG encryption to safely store and retrieve credentials, ensuring that sensitive information is shielded from unauthorized access. Lab managers can access a centralized password store without the risks associated with sending passwords through email or messaging platforms like Slack. This system simplifies secure password sharing among authorized users and enhances overall security in managing sensitive access within a lab environment.

**Internal documentation**: To manage the different tasks related to IT management, IT documentation should be kept in a digital notebook; again, we use GitHub here, keeping the IT notes in a section of our intranet. We also keep a separate section of sensitive information is present (eg: IP addresses, sensitive messages) in a the repository can be made private for the lab; this does not build to a public website, but it does use the same markdown format that our public websites do, and those notes render cleanly and legibly directly on GitHub.

---

[10] https://www.netdata.cloud/
[11] https://archlinux.org/packages/extra/any/pass/

## Student's lab notes

We don't have paper lab notes. Only digital. Typically, a student would have a Google Doc (GDoc) as a lab note called 'progress report'. Permissions are easy to modify. Eg: that way, if someone becomes a co-supervisor, it is easy to share that lab note. Lab note is organized in anti-chronological order, per meeting, with the date of the meeting as the headings. Note that the date of the headings, as well as for naming documents, should always follow the standard: YYYY-MM-DD. Also see **Figure 4**.

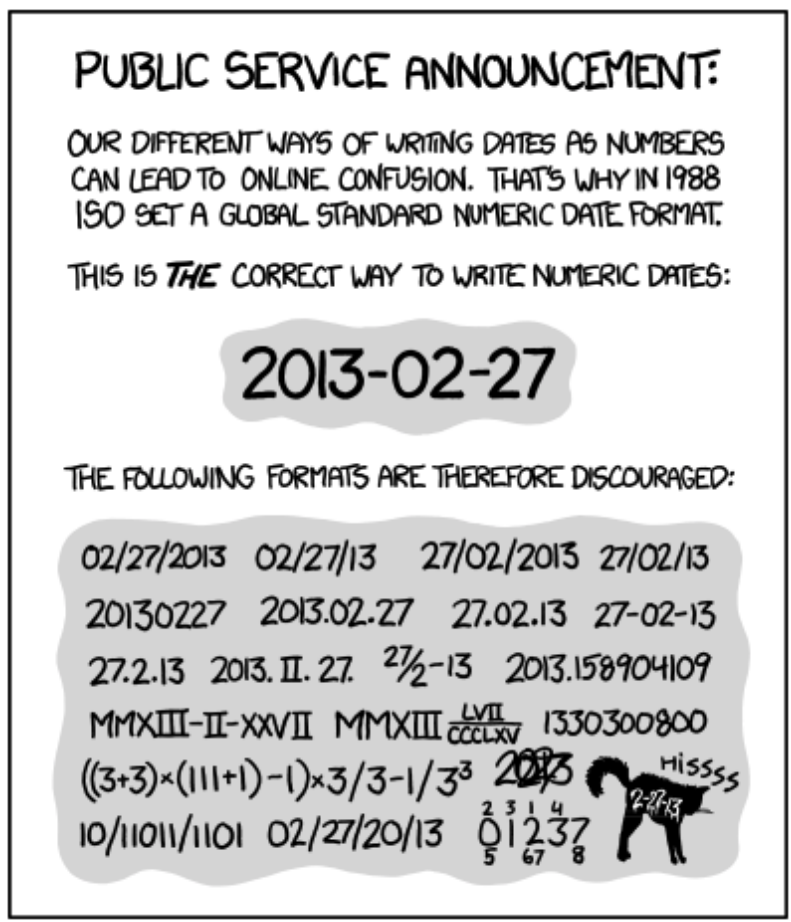

**Figure 4**. **How to properly format dates?** Especially when naming electronic files. Source: https://xkcd.com/1179/

## Project management

Every project has a shared GDoc associated with it. It follows the same organization as for student's lab notes. Agenda and minutes are organized per meeting. At the top of the GDoc, there is a link to the GitHub repository of the project, calendar to the meetings with a Zoom link for the meetings. To make it easier to have external collaborators participate in our lab meetings, those lab meeting agenda are usually public. See e.g. the one for Spinal Cord Toolbox[12].

In all projects, whether this is a software dev project, or an MR physics project, we try to have a GitHub repository associated with it. Then, students use GitHub issues as a way to ask questions about the project, document their experiments. GitHub issues is a fantastic way to keep track of discussions, eg: we can cross-reference the code that was used to generate some results, we can add images, we can format text and code with markdown. To ensure transparency and reproducibility, when presenting results of an experiment, always specify the version of the code that was used (SHA), and the version of the data (SHA from git-annex/Datalad). Ideally, the analysis script should produce log files that include these versioning information (also see YODA framework[13])

Issues are also a great way to initiate project ideas, and assign labels to it (eg: 'good internship project', or 'good first issue')-- then, once a student is assigned to the project, GitHub has a tag 'assigned to XX'. Issues can also be referenced in a Kanban project managing board. The best resource to get started with leveraging the power of github's issues to manage code and project is this link: https://docs.github.com/en/issues.

## Open source coding

In the dynamic and collaborative environment of a research lab, adopting open source coding practices, especially through platforms like GitHub and GitLab, offers a multitude of benefits that enhance both the development process and the educational experience of students. One significant advantage is the facilitation of code review via pull requests, a practice that not only ensures the quality and maintainability of the codebase but also teaches students the importance of writing clear, readable, and well-documented code—a skill highly valued in the industry. This process promotes a culture of collaborative learning and peer feedback, crucial for the professional growth of budding researchers and developers.

Moreover, the integration of continuous integration (CI) tools within these platforms allows for automatic code quality checks against standards like PEP8 in Python, and the use of linters, as well as the execution of unit and integration tests. This automated feedback loop not only helps in maintaining a high standard of code quality but also in ensuring the functionality and reliability of the software being developed, crucial for the credibility of research outputs.

---

[12] https://docs.google.com/document/d/1ItApJQfajO2lRzOU2yenWbeRg6alfsdut3J4AVVdo78
[13] https://handbook.datalad.org/en/latest/basics/101-127-yoda.html

The open-source model further amplifies these benefits by fostering a transparent, inclusive, and collaborative ecosystem. Coding in the open, as highlighted in discussions by government digital services, encourages innovation, facilitates the sharing of knowledge and resources, and accelerates problem-solving by leveraging the collective expertise of a wider community. It also significantly enhances reproducibility and transparency in research, allowing findings to be more easily validated and built upon by others in the field. Additionally, engaging with the open-source community can provide valuable feedback from a diverse array of users and contributors, leading to more robust, versatile, and user-friendly research tools.

In sum, the adoption of open source coding practices within research labs not only improves the quality of software development and research outcomes but also prepares students for successful careers in both academia and industry by instilling best practices in coding, collaboration, and open science.

## Writing articles

**Collaborative writing**: Exception to open source: we use GDoc and Paperpile[14]. Some students use Latex-based Overleaf[15] (mostly ML students). These technologies enable researchers to collaborate simultaneously on the same version of the manuscript, add comments, see who wrote the comments (if they are logged in with their google account), and tag people on comments. I find this makes article writing very efficient. Also note there is a convenient "version history" feature in GDocs, that lists the changes and the authors of those changes. Example of large article-writing collaboration using that approach: [5] (92 co-authors), [6] (57 co-authors). The wrong way to do it (in my humble opinion) is for one author to write the manuscript on a local file (eg: Microsoft Word), and then send the article to all co-authors via email. Working on multiple 'unsynchronized' versions of a document rather than using a single, cloud-based document with version control is problematic for several reasons:

- It leads to confusion and inefficiencies as team members may not be aware of the most current version, resulting in redundant efforts or the incorporation of outdated information.
- This approach often results in the circulation of dozens of emails, each containing a different version of the document. Particularly when figures are large, attachments can exceed 10MB, leading to some co-authors not receiving emails due to size restrictions or spam filters. This not only inflates email quotas but also duplicates physical data on the cloud, an action that is increasingly recognized as irresponsible in the context of climate change concerns.
- Furthermore, the task of merging all these disparate changes is fraught with challenges. It is prone to human error and typically falls to a single individual, making the process opaque to other co-authors who are unable to track changes or understand decisions made during the merging process. This lack of transparency can lead to miscommunications and missed opportunities for collaboration and improvement.

---

[14] https://paperpile.com/
[15] https://overleaf.com/

- This method often results in a chaotic assortment of file names, creating confusion and making it difficult to identify the most recent or relevant version of the document. This situation is humorously exemplified in **Figure 5**.

In contrast, a single document managed on a cloud platform with version control streamlines collaboration by ensuring all team members have access to the latest version, facilitates the tracking of revisions, and simplifies the process of integrating contributions. This approach not only enhances productivity but also improves the accuracy and integrity of the document.

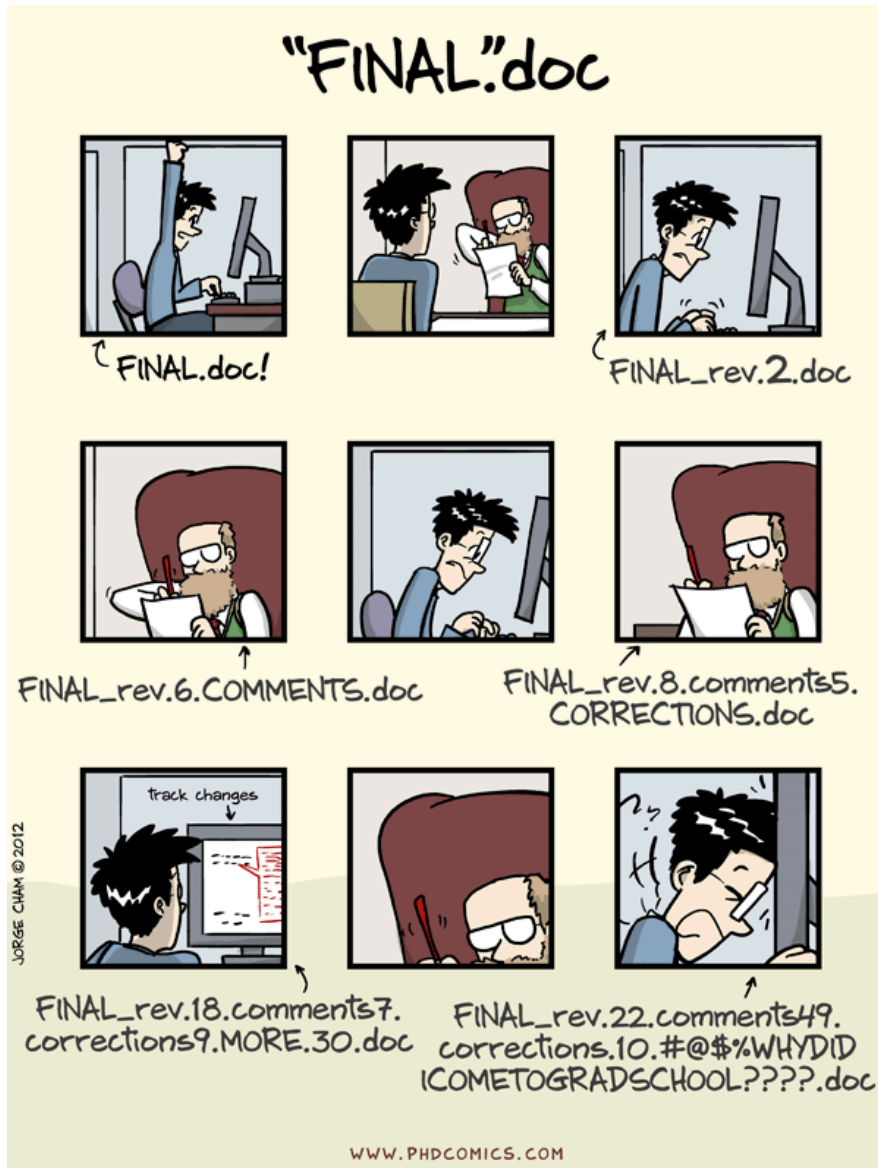

**Figure 5**. Why is it a bad idea to work on multiple 'unsynced' versions of a document, versus a single document on the cloud one that has version control? Source: https://phdcomics.com/

**Promoting reproducible science:** In spirit with the concept of reproducible science, the whole paper should, in principle, be self-reproduced with the data and the code available– at least the results section with the figures. To make sure we don't end up in a situation like on **Figure 1**, the entire code should be able to reproduce *exactly* what is on the publication. The best way to do it, is to submit the code alongside the paper, or even better, to have the figure of the paper being generated on the fly when opening the paper. This approach is made possible with the NeuroLibre reproducible preprints[16], and it is a nice way moving forward, not only for publication, but for any ongoing projects.

One last thing on that topic: When publishing an article/abstract where the code was put on Github, add a link to the GitHub/GitLab release, not to the landing page of the repository which by default shows the main/master git branch. A release points to code that is frozen in time (ie: when paper was published), whereas the main/master branch points to code that can evolve (this is the *raison d'être* of git: to track evolution).

## Communication

**Rapid communication:** Exception to open source: we use Slack. We tried mattermost. It was too cumbersome to deploy and to maintain the server locally. But the biggest hurdle was adoption: lab members and collaborators already had Slack installed and used it. It was too much to ask to use another app in their daily routine. We ended up with a hybrid mode, with some people using Slack, and a minority using mattermost. It was a disaster. We stopped it and went back to full Slack with a paid license (about 800 USD per year with ~244 members and academic discount).

**Forums:** To manage software projects, or communities, an excellent open-source technology for setting up forums is 'Discourse' (https://www.discourse.org/). There are excellent tutorials out there to explain step-by-step how to host, deploy and manage such forums.

**Calendars:** The lab also makes good use of shared calendars. It's another exception to open source since it's GCal, but it should be supported on any other calendar software. Example of useful shared calendars:

- **Lab meeting calendar**. It's nice to know what everyone is up to, avoids scheduling conflicts, central place to put room number, link to minutes, link to zoom room.
- **Vacation/conference calendar**. Also good for having a general sense of the lab. Can also contain submission deadlines as a reminder.
- **GPU calendar**. It's a nice, low-friction way of managing shared resources in the lab.

---

[16] https://neurolibre.org/

# Conclusion

The philosophy behind using open source software and exposing students to them for managing a lab, is to encourage the development of such software (by contributing to them), and to encourage 'by the example' other labs to use them in their daily practice. In the era of transparent and reproducible science, it is elusive to try to make science reproducible by ONLY focusing the efforts at the time of publication (which represents 1% of the scientific contribution)-- many things have happened before: meetings, informal discussions between colleagues, study design, data collection, code design, data analysis. All this information should also be Findable, Accessible, Interoperable and Reusable (these are the four concepts behind the FAIR principles).

We should note, however, that even if all these tools are useful, some of them require good IT skills (e.g. ansible, git-annex). There is a strategic trade-off between the time it takes to invest on a new technology, versus the time this technology will save. I like to illustrate this point with **Figure 6**. Also include the fact that these tools not only save time, but expose students to excellent practices, and ensure some level of transparency and reproducibility.

To sum up, here are the benefits of using open source technologies to manage a lab and do science:

- Gain useful skills (git, code review, etc.)
- Your research is transparent and can be reproduced
- People use your stuff (rewarding, boosts h-index)
- People give you feedback ➔ your project gets better
- Publicly funding projects should serve society

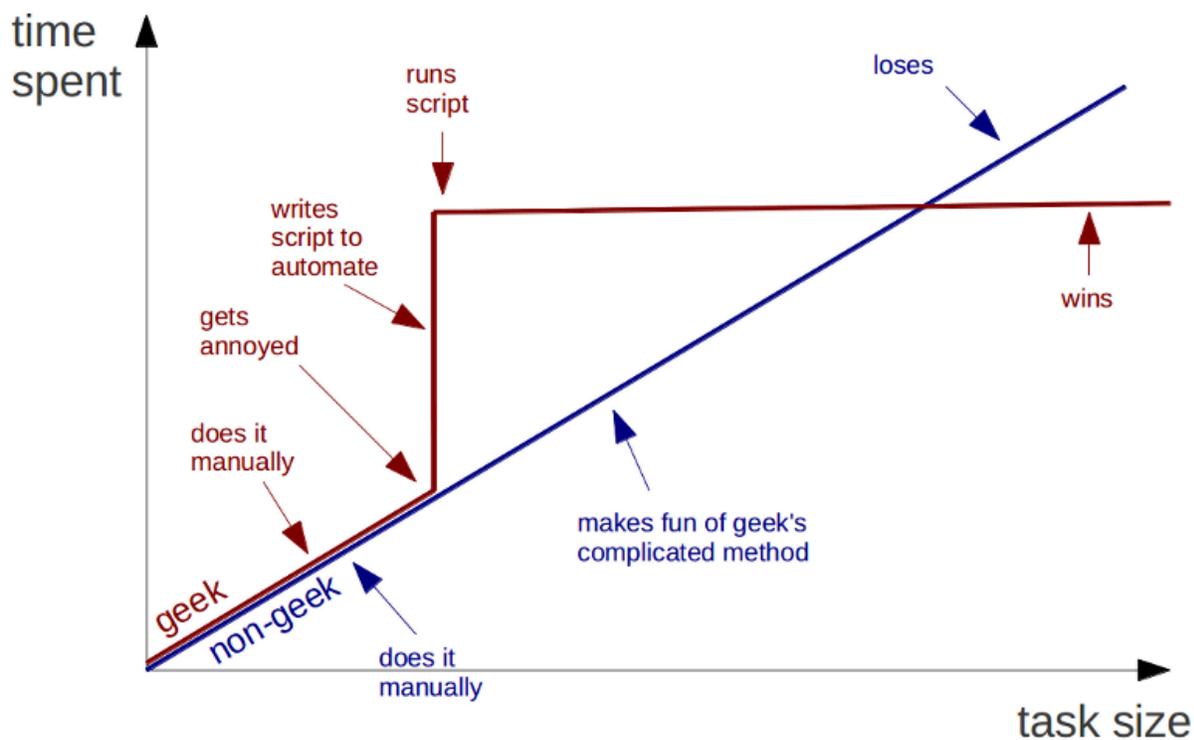

**Figure 6**. Is it worth spending time developing automated procedures? This chart can answer that. Source: https://i.imgur.com/Q8kV8.png

## Acknowledgements

I would like to thank Sandrine Bédard, Naga Karthik Enamundram, Jan Valošek, Armand Collin, Mathieu Guay-Paquet and Nick Guenther for contributing to this document.